# Interlayer ferroelectric polarization modulated anomalous Hall effects in four-layer MnBi$_2$Te$_4$ antiferromagnets

Ziyu Niu[1], Xiang-Long Yu[3], Dingfu Shao[4], Xixiang Jing[1], Defeng Hou[1], Xuhong Li[1], Jing Sun[1], Junqin Shi, Xiaoli Fan[1,2]* and Tengfei Cao[1,2]*

[1]*Research Center for Advanced Lubrication and Sealing Materials, School of Materials Science and Engineering, Northwestern Polytechnical University, Xi'an, Shaanxi 710072, P.R. China.*
[2]*State Key Laboratory of Solidification Processing, Northwestern Polytechnical University, Xi'an, Shaanxi 710072, PR China.*
[3]*Shenzhen Institute for Quantum Science and Engineering, Southern University of Science and Technology, Shenzhen 518055, China*
[4]*International Quantum Academy, Shenzhen 518048, China.*
[5]*Key Laboratory of Materials Physics, Institute of Solid State Physics, HFIPS, Chinese Academy of Sciences, Hefei 230031, China.*
**Email:** xlfan@nwpu.edu.cn, Tengfei.Cao@nwpu.edu.cn

## Abstract

Van der Waals (vdW) assembly could efficiently modulate the symmetry of two-dimensional (2D) materials that ultimately governs their physical properties. Of particular interest is the ferroelectric polarization being introduced by proper vdW assembly that enables the realization of novel electronic, magnetic and transport properties of 2D materials. Four-layer antiferromagnetic MnBi$_2$Te$_4$ ($F$-MBT) offers an excellent platform to explore ferroelectric polarization effects on magnetic order and topological transport properties of nanomaterials. Here, by applying symmetry analyses and density-functional-theory calculations, the ferroelectric interface effects on magnetic order, anomalous Hall effect (AHE) or even quantum AHE (QAHE) on the $F$-MBT are analyzed. Interlayer ferroelectric polarization in $F$-MBT efficiently violates the $\hat{P}\hat{T}$ symmetry (the combination symmetry of central inversion ($\hat{P}$) and time reverse ($\hat{T}$)) of the $F$-MBT by conferring magnetoelectric couplings, and stabilizes a specific antiferromagnetic order encompassing a ferromagnetic interface in the $F$-MBT. We predict that engineering an interlayer polarization in the top or bottom interface of $F$-MBT allows converting $F$-MBT from a trivial insulator to a Chern insulator. The switching of ferroelectric polarization at the middle interfaces results in a direction reversal of the quantum anomalous Hall current. Additionally, the interlayer polarization of the top and bottom interfaces can be aligned in the same direction, and the switching of polarization direction also reverses the direction of anomalous Hall currents. Overall, our work highlights the occurrence of quantum-transport phenomena in 2D vdW four-layer antiferromagnets through vdW assembly. These phenomena are absent in the bulk or thin-film in bulk-like stacking forms of MnBi$_2$Te$_4$.

## I. INTRODUCTION

The manipulation of symmetry can yield distinct phases of matter and, thus, unique material properties[1]. The time-reversal symmetry ($\hat{T}$) breaking directly impacts spin polarizations and leads to the generation of spontaneous magnetic orders, forming the foundation for spintronic devices. Magnetic field, either external or intrinsic one, can be applied to break the time-reversal symmetry ($\hat{T}$) of the system and induces spin polarizations. However, large amounts of electric current are typically required to generate effective magnetic field, which could cause significant energy losses. Additionally, for fully compensated antiferromagnets, magnetic field cannot efficiently control their magnetic orders and poses significant limitations for applications of antiferromagnets in electronic devices[2-8].

It is anticipated that electric voltage-controlled magnetization could offer noteworthy benefits for non-volatile, low-power spintronic devices[9,10]. On the occasion of the centenary of ferroelectricity discovery, Wu *et al*. predicted interlayer ferroelectricity for the first time in two-dimensional van der Waals (vdW) materials[11]. This theoretical breakthrough has the potential to significantly expand our understanding of ferroelectric properties and drive groundbreaking technological advancements in thin-film and multilayered nanostructured materials. Interlayer ferroelectricity has been confirmed by a series of vdW assembly experiments[12-18]. Ferroelectric polarization induced by interlayer stacking, even relatively weak (~0.01μC/cm$^2$) compared with perovskites (28.6 μC/cm$^2$ for PbTiO$_3$)[19] or binary metal-oxide (51.39 μC/cm$^2$ for HfO$_2$)[20], enables novel electronic and transport properties that were previously unattainable in bulk materials or vdW films in bulk-like stacking patterns[17,18,21-23]. It is because interlayer polarization could efficiently break intrinsic symmetries of crystals, and lifts symmetry restrictions on electronic properties of different crystals. For example, by introducing interlayer polarization, the $\hat{P}\hat{T}$ symmetry of the system is broken, which removes Kramers degeneracies of band structures and greatly impacts their electronic transport properties. Furthermore, the interlayer ferroelectricity generated through vdW assembly not only expands the range of ferroelectric materials but also provides an efficient means to explore novel properties that were previously unattainable in bulk materials or bulk-like stacking films[18].

Among all 2D vdW materials, the introduction of interlayer polarization into the antiferromagnetic topological family MB$_2$T$_4$ (M=Ti, V, Cr, Mn, Fe, Co, Ni, Eu; B=Bi, Sb; C=Te, Se) stands out as the most fascinating. The magnetic, ferroelectric, and topological transport properties in MB$_2$T$_4$ family can potentially interact in interesting ways, leading to a direct impact on the occurrence of quantum phenomena such as anomalous hall effect (AHE) or even quantum AHE (QAHE)[2].

MnBi$_2$Te$_4$, serves as a prototype member of the MB$_2$T$_4$ family, which exhibits *A*-type antiferromagnetic order[4], and has been successfully synthesized in experiments[24]. The interaction between magnetic ordering and topological states can yield a range of quantum phenomena[25-28]. Specifically, the emergence of quantum states in MnBi$_2$Te$_4$ thin films is strictly dependent on its layer numbers *N*. When *N* is even, there is no detectable anomalous Hall conductance (AHC) and it is predicted to be an axion insulator[29,30]; when *N* is odd, it is able to show a plateau of quantum anomalous Hall conductance of $e^2/h$ around the Fermi energy level, associating with QAHE[5]. The phenomenon can be explained in terms of different magnetic space group in even or odd layers of MnBi$_2$Te$_4$.

Four-layer MnBi$_2$Te$_4$ (*F*-MBT) structures are special existing states of MnBi$_2$Te$_4$ thin films. On the one hand, they inherit the $\hat{P}\hat{T}$ symmetry of the parent bulk phase; On the other hand, they acquire surface structures that are not available in the bulk phases. Furthermore, compared with bilayer MnBi$_2$Te$_4$ films with $\hat{P}\hat{T}$ symmetry, *F*-MBT films has three interlayer spaces. Interlayer polarization could be induced and aligned along different directions, and introduces different coupling effects among electronic dipoles, magnetic orders as well as topological states. Therefore, *F*-MBT provides an ideal basis for exploring the arrangement of interlayer polarizations, magnetoelectric couplings and topological states in MnBi$_2$Te$_4$ under different polarization arrangements. Based on ideas that ferroelectric polarization breaks intrinsic symmetry of *F*-MBT[31], we expect that the arranging of interlayer polarization at different locations will produce attractive physical correlations between ferroelectricity, magnetic order, topological transport properties and surface states.

Here, by combining symmetry analysis and first-principles calculations, we show that interlayer ferroelectric polarization could efficiently induce AHE or even QAHE in *F*-MBT films. Specifically, because of $\hat{P}\hat{T}$ symmetry breaking of the *F*-MBT, non-zero AHC can always be obtained. Under the influence of interlayer ferroelectric polarization, it is possible to stabilize a unique antiferromagnetic order that includes a ferromagnetic interface and maintains a zero net magnetic moment. Our study demonstrates reversible quantized topological transport properties in magnetically fully compensated *F*-MBT relying solely on the interlayer polarization, which has not been reported in previous work, and our predictions propose new avenues for efficiently controlling magnetic symmetry, consequently influencing spin-dependent topological transport properties, without the need for an external magnetic field.

## II. METHODS

We perform all density functional theory (DFT) calculations by applying the Vienna Ab initio Simulation Package (VASP)[32]. The exchange and correlation effects are treated within the generalized gradient approximation (GGA)[33]. The GGA+U functional with $U_{eff}$ = 4 eV on Mn 3d orbitals is applied in all calculations[34,35]. The value of U has been tested by Li *et al*[4]. For MnBi$_2$Te$_4$ and it was found that $U$ = 4 eV provides the band structures that match well with the experiments. Moreover, the U = 4 eV has been widely utilized in numerous theoretical studies to investigate the quantum transport properties of MnBi$_2$Te$_4$[31,36,37]. The plane-wave energy cutoff is set to 500 eV in all calculations. A 15 Å vacuum region was used between adjacent plates to avoid the interaction between neighboring periodic images. The vdW corrections, as parameterized within the semiempirical DFT-D3 method[38], are used in the calculations. The systems are fully relaxed until the residual force on each atom is less than 0.01 eV/Å. Electronic minimization is performed with a tolerance of $10^{-7}$ eV. Unless mentioned in the text, spin-orbit coupling (SOC) is always included in the calculations of the electronic and transport properties. The out-of-plane electric polarization of polar-stacked MnBi$_2$Te$_4$ films is calculated using the dipole correction method[39]. The maximally-localized Wannier functions are used to obtain the tight-binding Hamiltonians within the Wannier90 code[40]. The Berry curvature, the anomalous Hall conductivity and the Chern number are calculated using the Wanniertools code[41].

## III. RESULTS AND DISCUSSION

Fig. 1(a) presents a four-layer bulk-like stacked vdW magnetic MnBi$_2$Te$_4$ film (*FB*-MBT), where each MnBi$_2$Te$_4$ layer is composed of one MnTe layer and one Bi$_2$Te$_3$ layer, and the MnTe layer is embed into Bi$_2$Te$_3$ matrix. Being similar to its parent bulk counterpart, the *FB*-MBT exhibits an out-of-plane A-type antiferromagnetic order[28] (Table. S1 shows the relative energies of *FB*-MBT in different magnetic orders), and it also inherits the intrinsic $\hat{P}\hat{T}$ symmetry of the antiferromagnetic MnBi$_2$Te$_4$ bulk phase. Because of $\hat{P}\hat{T}$ symmetry restriction, there are no net magnetic moments in the *FB*-MBT, which prevents AHE or even QAHE that could be detected in the *FB*-MBT film. The $\hat{P}\hat{T}$ symmetry restriction on the AHE or the QAHE can be understood from the following intrinsic definition of the berry curvature ($\Omega(k)$) and anomalous hall conductivity ($\sigma_{xy}$),

$$\Omega(k) = -2Im\sum_{m\neq n}\frac{\langle n|\frac{\partial \hat{H}}{\partial k_x}|m\rangle\langle m|\frac{\partial \hat{H}}{\partial k_y}|n\rangle}{(E_{nk}-E_{mk})^2} \quad (1)$$

where $\hat{H}$ is the Hamiltonian of the system, and $E_{nk}$ is the energy of the $n_{th}$ band at wave vector $k$. $|m\rangle$, $|n\rangle$ are corresponding eigenstates of $\hat{H}$. The intrinsic anomalous hall conductivity ($\sigma_{xy}$) is calculated through the integration of $\Omega(k)$ in the whole Brillouin zone (BZ) (equation (2)):

$$\sigma_{xy} = -\frac{e^2}{h}\int_{BZ}\frac{d^2\kappa}{2\pi}\Omega(k) \quad (2)$$

From the definition of $\Omega(k)$ in equation (1), it can be obtained that $\Omega(k)$ is an odd function under the $\hat{P}\hat{T}$ symmetry operation ($\hat{P}\hat{T}\Omega(k) = -\Omega(-k)$), which makes $\Omega(k)$ zero and the integration of $\Omega(k)$ in the whole BZ zero. Therefore, the AHE or the QAHE cannot be detected in the *FB*-MBT.

The $\hat{P}\hat{T}$ symmetry constraint on the AHE or the QAHE can be eliminated by manipulating the stacking patterns of MBT layers[31,42], a possibility that can be explored through experimental vdW assembly. Here, based on the *FB*-MBT (Fig. 1(a)), we show that $\hat{P}\hat{T}$ symmetry can be broken in four-layer MnBi$_2$Te$_4$ through introducing interlayer polarization. To simplify, the structure of a four-layer MnBi$_2$Te$_4$ is depicted in Fig. 1(a), with interlayer polarization being induced into the interlayer space between the top two layers. The interlayer atoms that generate out-of-plane ferroelectric polarization are highlighted.

In order to introduce interlayer polarization into this top two layers, we initially accomplished a 180° rotation of the top layer to break the inversion symmetry of the system. This rotational adjustment leads to the formation of top two-layer MnBi$_2$Te$_4$ film in *AA*-stacking (Fig. 1(b)). Furthermore, such non-centrosymmetric *AA*-stacked two-layer MnBi$_2$Te$_4$ film is unstable. The top MnBi$_2$Te$_4$ layer can spontaneously slip along the in-plane **t**$_{//}$ (1 -1 0) direction of the film or -**t**$_{//}$ direction, as indicated by the arrows in Fig. 1(e), to achieve the most stable electric polar structures. For clarity and improved readability, the term "*FP*-MBT" is used to refer to the four-layer MnBi$_2$Te$_4$ films with polar interfaces. The interlayer polarization in MnBi$_2$Te$_4$ is obtained with the Bi atoms in the bottom layer being aligned with the Te atoms in the top layer (Fig. 1(c)), or Bi atoms in the top layer being aligned with the Te atoms in the bottom layer (Fig. 1(d)). Correspondingly, the direction of interlayer polarization points up ($P_u$) or points down ($P_d$), respectively. For all *FP*-MBT films containing interlayer polarization, the original $\hat{P}\hat{T}$ symmetry of *FB*-MBT has been removed, which greatly impacts their magnetic ground states as well as the related topological transport properties.

We illustrate that magnetic ground states of *FP*-MBT are determined by interlayer polarizations, and these magnetic ground states with the corresponding topological transport properties of *FP*-MBT films are depicted schematically in Fig. 2. For the *FP*-MBT, there are three interlayer spaces: the top interspace (*TI*), the middle one (*MI*) and the bottom one (*BI*), respectively. The interlayer polarization could be introduced into the *TI*, *MI*, *BI*, both *TI* and *MI*, both *TI* and *MI*, or even in all interlayer spaces through vdW assembly. For simplicity, for *FP*-MBT structure with interlayer polarization in *TI*, *MI*, both *TI* and *MI*, both *TI* and *BI* or in all interlayer spaces, is referred to as *TIP*, *MIP*, *TMIP*, *TBIP* or *AIP*, respectively. The interface polarization can point-up ($P_u$) or point-down ($P_d$), and is simply referenced as *TIP*-, *MIP*-, *TMIP*-, *TBI*P- or *AIP*-$P_u$ ($P_d$), correspondingly. We found that the magnetic ground state is closely connected to the interlayer polarizations in *FP*-MBT.

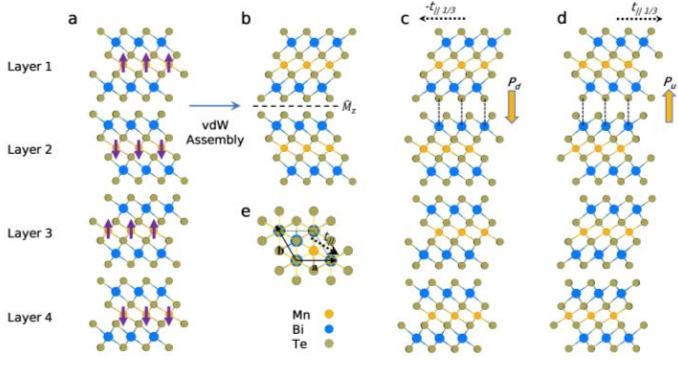

**Fig. 1** (**a, e**) Side and top views for atomic structures of four-layer MnBi$_2$Te$_4$ in bulk-like stacking. The purple arrows in panel (a) represent the direction of magnetic moments of Mn atoms. (**b**) A mirror-reflected ($\hat{M}_z$) bilayer structure was achieved by vdW assembly of layer1 and layer2 in the *FB*-MBT film without considering the bottom two-layers (a), leading to an unstable AĀ stacking configuration. (**c, d**) Schematic representation for the formation of a four-layered MBT film with an interlayer polarization $P_d$ (c) or $P_u$ (d) between layer1 and layer2 by sliding layer1 along the -$t_{//1/3}$ or $t_{//1/3}$ (e) direction within the plane from the AĀ stacking in (b).

For *FB*-MBT without interlayer polarization, the *A*-type up-down-up-down (↑↓↑↓) magnetic order is the magnetic ground state. For *TIP*-$P_u$/$P_d$ and *MIP*- $P_u$/$P_d$, our calculations show that the most stable magnetic order is (↑↓↓↑) (Fig. 2(a, b), Table S2, S3). For *TMIP* with two polar interfaces, the magnetic order could be (↑↓↑↓) or (↑↓↓↑) (Fig 2(c), Table S4), depending on the direction of interlayer polarization, and for *TBIP* with two polar interfaces, the magnetic ground state is still *A*-type antiferromagnetic order (↑↓↑↓) (Table S6). The topological transport properties of polar *FP*-MBT films are different from that of *FB*-MBT, which are directly determined by interlayer polarization and magnetic orders. Firstly, interlayer polarization has lifted the $\hat{P}\hat{T}\Omega(k) = -\Omega(k)$ symmetry constraint on the *FB*-MBT, and it results in a nonzero berry curvature and none-zero AHC in all *FP*-MBT films. Furthermore, we find that the (↑↓↓↑) magnetic order has the lowest energy in *TIP*, *TMIP* and *MIP* (Fig. 2(a-c), Table S2-4) among all magnetic states. Finally, as it is schematically given in Fig. 2 (a-c), even the net magnetic moment for (↑↓↓↑) or (↓↑↑↓) magnetic order in four-layer MnBi$_2$Te$_4$ is still zero, the ferromagnetic coupling in the middle interface could also induce quantum anomalous Hall conductivities (QAHC) in *FP*-MBT, being similar to MnBi$_2$Te$_4$ films having net magnetic moment.

In the following paragraphs, we will specifically discuss interlayer polarization effects on magnetic order and topological transport properties of *FP*-MBT films.

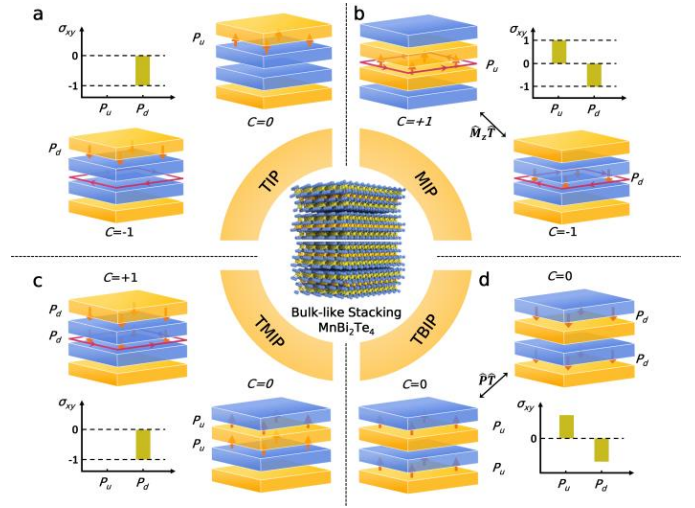

**Fig. 2** (**a**) Bottom left and top right: schematic structure of polar stacked *TIP FP*-MBT with electric interlayer polarization directions in top interface of *FP*-MBT $P_d$ and $P_u$, respectively. The schematic yellow layer represents that the magnetic moment of Mn atoms in the whole layer pointing up (↑), and the blue one indicates that all magnetic moments point down (↓). The orange arrows in the interlayer space show the direction of interlayer polarization. Upper left: the AHC corresponding to the stacking pattern of the $P_d$ and $P_u$ at the Fermi energy level. (**b**) Bottom right and top left: the schematic diagram of the *MIP*-$P_d$/$P_u$ stacking structure, where the electric interlayer polarization exists in the middle interfaces of the *FP*-MBT. Here the $P_d$ and $P_u$ states are connected by $\hat{M}z\hat{T}$ symmetry (the combination of in-plane mirror ($\hat{M}z$) and time reversal ($\hat{T}$) symmetry). In the pattern of *MIP*-$P_d$, the Chern number is -1 and the Chern number is +1 for *MIP-P$_u$*, which is confirmed by the AHC at the Fermi energy level in the upper right corner. (**c**) Bottom right and top left: schematic conformation of a polarized *FP*-MBT films stacking mode between two adjacent interlayers with two identical orientations (*TMIP-P$_u$/P$_d$*), whose Chen numbers is 0 and 1, respectively, as confirmed by the AHC at the Fermi energy level (bottom left). (**d**) Bottom left and top right: Schematic representation of *TBIP-P$_u$/P$_d$* stacking models that retain the *A*-type antiferromagnetic order in the bulk phase and are associated by $\hat{P}\hat{T}$ symmetry operation. Schematic of the AHC at *E-E$_f$* = 0.05eV, which implies that *TBIP-P$_u$/P$_d$* remain topologically trivial insulators in these stacking patterns. The specific atomic structures of all the stacking pattern mentioned above can be found in the Fig. 1(c, d) and in supplementary materials Fig. S1.

Fig. 3(a) shows one example of the energy profile and interlayer polarizations of *FP*-MBT system, where top layer slides along the [1$\bar{1}$0] direction with respect to bottom three bulk-like stacked MnBi$_2$Te$_4$ films. The fractional value of **0** corresponds to the mirror-symmetric ($\hat{M}_z$) stacked layer1 and layer2 in *FP*-MBT. It is a high energy unstable stacking pattern. The structure at the fractional value of **1** is the same with that of **0**, which indicates that the top layer has shifted the whole crystal periodicity of the system. The two energy minima appear at fraction of 0.33 and 0.66 of the whole sliding paths, corresponding to the structure of TIP with interlayer polarization pointing up (*TIP-P$_u$*), or down (*TIP-P$_d$*), respectively. In these structures, the Bi/Te or Te/Bi at the interface of layer1 or layer2 aligns along the *Z* direction, generating electronic dipole moment pointing from Te to Bi atoms. The calculated electric polarization is approximately 0.01μC/cm$^2$, being almost the same to previous results on polarization stacking in bilayer MnBi$_2$Te$_4$ films[31].

According to the theory of super-exchange interaction between magnetic ions in the systems, super-exchange interaction occurs when an intermediate non-magnetic atom or ion mediates the exchange coupling between two magnetic ions. Here, intermediate non-magnetic atoms are Te and Bi atoms, and magnetic ones are Mn atoms. The interlayer magnetic coupling occurs when indirect electron hopping between the *d* orbitals of a magnetic element is mediated by connected *p* orbitals of Bi and Te atoms.

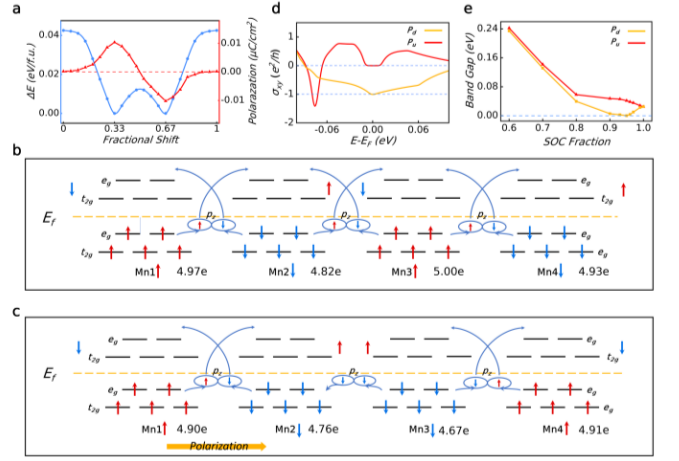

**Fig**. 3 (**a**) Total energy and out-of-plane electric polarization of a polar MnBi$_2$Te$_4$ bilayer when sliding the top monolayer with respect to the bottom along the [1$\bar{1}$0] direction. The two minima in energy profile correspond to polarization in up ***P$_u$*** and down ***P$_d$*** direction. (**b**) Schematic diagram of the antiferromagnetic super-exchange interactions for Mn atoms in four-layer bulk-like stacked MnBi$_2$Te$_4$ film. In (b, c), the arrows denote the direction of spin polarization of Mn atoms. The light blue curves with an arrow denote hopping paths. (**c**) The ferromagnetic coupling between Mn2 and Mn3 in four-layer MnBi$_2$Te$_4$ containing one interlayer polarization between Mn1 layer and Mn2 layer. (**d**) Anomalous Hall conductance $\sigma_{xy}$ of the *TIP-P$_d$/P$_u$* as a function of electron energy. (**e**) Variation of the band gap for polarization ***P$_u$*** and ***P$_d$*** of *TIP-P$_d$/P$_u$* as a function of fractions of the actual SOC.

In each layer of MnBi$_2$Te$_4$, the Mn atom is surrounded by Te atoms in a slightly distorted octahedral configuration. This configuration results in the splitting of *d*-orbitals into a triple-generating *t$_{2g}$* state and a double-generating *e$_g$* state. In MnBi$_2$Te$_4$ bulk phase, or MnBi$_2$Te$_4$ films in bulk-like stacking patterns, the electronic occupation number of each Mn atom is around 5$e$ (Fig. 3(b)), and there are few charge-transfer from $d$ orbitals of Mn atom to its surrounding Te atoms, the $e_g$-$p$-$e_g$ super-exchange (Fig. 3(b)) interaction plays a dominant role in determining the interlayer magnetic coupling (IMC) of MnBi$_2$Te$_4$, and the interlayer magnetic order is antiferromagnetic one.

For *FP*-MBT films containing electronic interlayer polarizations (Fig. 3(c)), our results show that the $d$ orbital occupation of Mn is efficiently modulated by the interlayer electric polarization. Here, we take *TIP-P$_d$* as an example and for clarification, the Mn atoms in four-layer MnBi$_2$Te$_4$ film from top layer to bottom one are labeled as Mn1, Mn2, Mn3 and Mn4, respectively. Our simulations show that their occupation number are 4.90e, 4.76e, 4.67e and 4.91e, respectively. It shows that, because of interlayer polarization, there are relatively large charge transfer from $d$ orbitals of Mn atom to $p$ orbitals of Te or Bi in middle two layers (the Mn2 layer and Mn3 layer) of *FP*-MBT, and at the same time the magnetic polarization is also partially forwarded from Mn to Te and Be atom in each corresponding layer. Therefore, the magnetic coupling between the Mn2 layer and Mn3 layer is the ferromagnetic one (Fig. 3 (c)). This specific ↑↓↓↑ magnetic order results directly from different occupations of Mn in each MnBi$_2$Te$_4$ layer. For example, the $d$ orbital occupation for both Mn1 and Mn4 is around 5e, being the same to the structure in bulk-like stacking, the $e_g$-$p$-$e_g$ super-exchange dominates interlayer magnetic interaction between Mn1-Mn2 and Mn3-Mn4, so both top and bottom interlayer space adopts antiferromagnetic couplings. The electron occupation of $d$ orbitals of Mn2 and Mn3 are 4.76e, 4.67e, and part of electrons in $d$ orbitals of Mn are shifted to Te and Bi atoms, which make the interlayer ferromagnetic couplings more favorable. In Fig. S5, we further calculated the energy difference between ↑↓↑↓ and ↑↓↓↑ order during the sliding process. It can be observed that with the increasing interlayer polarization strength, the ↑↓↓↑ magnetic order is gradually stabilized.

We further calculated the AHC (Fig. 3(d)) in *TIP* four films, and found that for both $P_u$ and $P_d$ *TIP*, they exhibit non-zero AHCs due to the breaking of $\hat{P}\hat{T}$ symmetry. However, there existed a noticeable distinction in the AHC around the Fermi level. Specifically, the $P_d$ state displays a quantized anomalous Hall conductance plateau, but the $P_u$ state shows a zero plateau. Moreover, the Chen number of the $P_d$ state is calculated to be 1, but that of $P_u$ is 0. All these prove that *TIP-P$_d$* state is a magnetic topological insulator, but *TIP-P$_u$* state is a trivial magnetic insulator. Furthermore, we have calculated the variation of the band gap with respect to the fractional strength of SOC. It can be observed in Fig. 3(e) and Fig. S2 that, for the *TIP-$P_u$*, its band gap gradually decreases with the increase of SOC strength, until it reaches the minimum bandgap of 24.5 meV at SOC fraction of **1**. There is no energy band inversion here, and proves a trivial band structure. In contrast, for the *TIP-P$_d$* state, its band gap first decreases with increasing SOC strength, completely closes at SOC=**0.95** (red curve in Fig. 3(e), Fig. S3), and then the bandgap reopens and increases in width until it reaches 28.8 meV at SOC=**1**. It proves that there is a band inversion in the energy band of *TIP-P$_d$* that possesses a nontrivial band structure, and produces quantized transport properties. The calculated berry curvatures for *TIP-P$_u$* and *TIP-P$_d$* are given in Fig. S4(a, b), which show that, for both $P_u$ and $P_d$ states, there is a sharp peak near the Γ point, while the berry curvature is nearly zero at other points. It indicates that the anomalous Hall conductivities in *TIP* are mainly contributed by the electronic states nearby the *Γ* point. Moreover, surface chiral edge state of *TIP-P$_d$* in Fig. S4 (d) also clearly shows that there is one single surface state crossing the bandgap at the Γ point, which is consist with the fact that *TIP-P$_d$* is a Chern insulator with Chen number equal 1. For *TIP-P$_u$* (Fig

S4 (c)), on the other hand, there are no chiral surface states could be observed and indicates normal semiconductor characters.

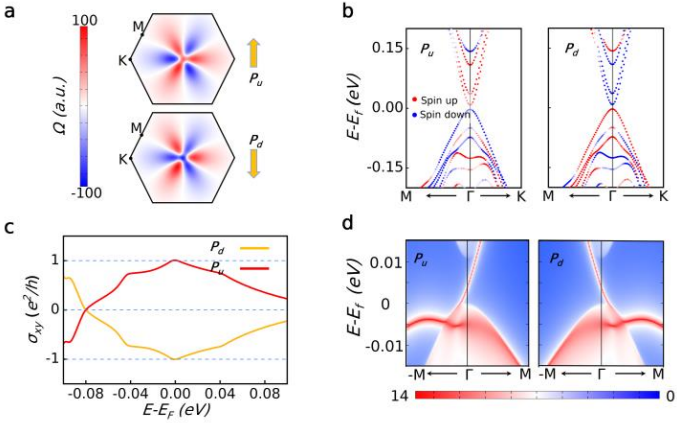

**Fig.** 4 (**a**) Berry curvature of the polar *MIP FP*-MBT in the 2D Brillouin zone calculated at the $E-E_f = 0.02$eV for polarization pointing-up (top) and -down (bottom). (**b**) Energy bands near the $\Gamma$ point of the *MIP F*-MBT with $P_u$ and $P_d$ interface polarizations. The red region represents a dominant spin-*up* configuration, while the blue area indicates a dominant spin-*down* one. (**c**) Anomalous Hall conductance $\sigma_{xy}$ of the *MIP FP*-MBT as a function of electron energy for polarization $P_u$ and $P_d$. (**d**) Surface band structures near the $\Gamma$ point of the *MIP FP*-MBT film for $P_u$ and $P_d$ demonstrate the appearance of the topologically protected edge states at the Fermi level.

Apart from the *TIP*, we further analyze interlayer polarization effects on the *MIP*, where the interlayer polarization is induced between layer2 and layer3 (Fig. S1(a, b)). Bing similar to *TIP*, the *MIP* could also have interlayer polarization pointing-up ($P_u$) or -down ($P_d$), and these two states can be switched by crossing an energy barrier of 0.01 eV/f.u.. The breaking of $\hat{P}\hat{T}$ symmetry in the *MIP* also leads to nonzero berry curvature as well as nonzero anomalous hall conductivities. The calculated berry curvatures in the 2D Brillouin zone for $P_u$ and $P_d$ states of *MIP* are given in Fig. 4(a). It shows that, for both $P_u$ and $P_d$ polar states of *MIP*, the berry curvature exhibits floral patterns, where the petals exhibit an alternating arrangement of red (positive value) and blue (negative value) colors. Moreover, the symmetry analysis for *MIP* indicates that the $P_u$ and $P_d$ are connects by $\hat{M}_z\hat{T}$ symmetry. Under $\hat{M}_z\hat{T}$ symmetry, the berry curvature of $P_u$ state ($\Omega_u(k)$) and that of $P_d$ state ($\Omega_d(k)$) is connected by the $\hat{M}_z\hat{T}\Omega_u(k) = -\Omega_d(-k)$ symmetry relation. It ensures that the color of each petal (reflecting the sign of $\Omega(k)$) remains independent of polarization. However, at the $\Gamma$ point where $k=-k$, as the polarization direction changes from upward to downward, the berry curvature $\Omega(k)$ also changes from positive value (isosurface color in red) to negative one (color in blue). Based on the nonzero berry curvature, we calculated the AHC for *MIP* in $P_u$ and $P_d$ states. Our results revealed a quantized plateau for both $P_u$ and $P_d$ state around the Fermi level with $\sigma_{xy} = \pm e^2/h$, which signifies the presence of quantum AHC (QAHC) in *MIP* state (Fig. 4(c)). Furthermore, QAHC exhibits opposite sign when the ferroelectric polarization changes from the $P_u$ state and the $P_d$ one. The presence of QAHC is associated with nontrivial topological features of the system, as evidenced by the topologically protected edge states shown in Fig. 4(d). From the figure, it can be observed that the edge states of $P_u$ and $P_d$ exhibit chiral symmetry, indicating not only the presence of nontrivial topology in both systems but also the symmetry correlations between them. We also considered the electronic band structures for upward and downward polarization in the Fig4 (b). Since the inability to accurately distinguish spin directions when considering SOC. The depth of red or blue represents the degree of dominance for spin-*up* or spin-*down*, respectively. Due to $\hat{M}_z\hat{T}E^{\uparrow}(k) = E^{\downarrow}(-k)$, the reversal of spin characteristics with polarization direction, the colors of the bands are opposite. Moreover, *FP*-MBT with polarization stacking exhibits magnetoelectric multiferroicity, possessing both spontaneous polarization and ferromagnetic orders. Importantly, the QAHE is achieved with nearly zero uncompensated net magnetic moment (approximately 0.003μB/f.u.), and it holds significant implications for enabling quantum devices resilient to external interference in the future.

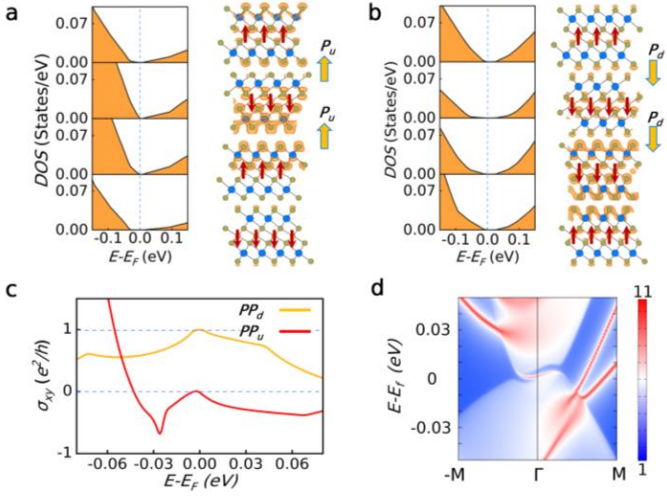

**Fig**. 5 *FP*-MBT film with interlayer polarizations in the top and middle interfaces (*TMIP*) and having polarization directions pointing-up *TMIP-P$_u$*(**a**) and pointing-down *TMIP-P$_d$* (**b**), and the corresponding layer-resolved density of states (DOS) as a function of energy (right panels). The real-space distributions of the charge density at the conduction band minimum are shown by oringin colored isosurfaces of 0.8% of their maxima. (**c**) Anomalous Hall conductivity as a function of electron energy for *TMIP-P$_u$P$_u$* and -*P$_d$P$_d$*. (**d**) Surface band structure of the *TMIP- P$_d$P$_d$* indicates the appearance of the topologically protected edge states at the Fermi level.

*FP*-MBT has three interlayer spaces, each capable of exhibiting interlayer polarization. Electric polarizations in different interlayer spaces can either accumulate or nullify each other. Firstly, we analyze the accumulation effects of interlayer polarizations on magnetic ground states and topological transport properties of *FP*-MBT. One interlayer polarization exists between layer1 and layer2, and the other one exists between layer2 and layer3 (see atomic structure in Fig. S1). Both of them point in the same direction, either up or down, and they are named *TMIP-P$_u$P$_u$* or *TMIP-P$_d$P$_d$*, respectively. Based on calculations of different magnetic orders, we find that *TMIP-P$_u$P$_u$* still maintains the bulk-like *A*-type antiferromagnetic order (↑↓↑↓). For the *TMIP-P$_d$P$_d$* states, the magnetic ground state is (↑↓↓↑), even the net magnetic moment is zero, there is ferromagnetic coupling between layer2 and layer3. These distinct magnetic orders indicate dissimilar magnetic transport properties of *FP*-MBT. We have calculated the AHC for both the upward and downward polarized structures (Fig 5(c)). In both cases, due to the breaking of $\hat{P}\hat{T}$ symmetry, the AHC above or below the band gap is non-zero. For the *TMIP-P$_d$P$_d$* state, there is a quantized $\sigma_{xy} = e^2/h$ plateau near the Fermi level (red curve) in the AHC curve, indicating a topological insulator with Chern number equal to 1, this is well supported by our calculations on the surface states in Fig. 5(d). Conversely, within the band gap region of the *TMIP-P$_u$P$_u$* structure, the AHC curve is zero (yellow curve) and the calculated Chen number is 0 proving a trivial insulator.

The AHC curves of *TMIP-P$_u$P$_u$*, *TMIP-P$_d$P$_d$* in the vicinity of the Fermi level exhibit different widths of zero plateau or the quantized $\sigma_{xy} = e^2/h$ plateau, reflecting different band edge bending effects. We further analyzed the layer-dependent density of states (DOS) demonstrated in Fig. 5 (a, b) to understand different interlayer polarization effects on band gap of *FP*-MBT. Our analysis revealed that the band gap of *TMIP-P$_u$P$_u$* is formed by the conduction band minimum (CBM) locating at layer2 or layer3 and valence band maximum (VBM) at layer2. For the *TMIP-P$_d$P$_d$* state, the width of the band gap is determined by the VBM at layer1 or layer4 and the CBM locates at layer3. Moreover, it can be observed in Fig. 5(a, b) that the band gap is mainly determined by the position of the CBM, because the energy levels of the VBM are almost the same for the *TMIP-P$_u$P$_u$* and the *P$_d$P$_d$* state. For the *TMIP-P$_u$P$_u$* state, the charge predominantly resides within layer2 and layer3, which are more proximate to the polarization interface. The interlayer polarization could exert a significant influence on the band edge of the CBM, resulting in a smaller band gap. For the *TMIP-$P_dP_d$* state, on the other hand, the CBM primarily localizes within layer3 and layer4, which has comparatively larger distance from the polarization interface than the case of *TMIP-$P_uP_u$*, consequently, the effect of polarization on the band edge is attenuated, and the band gap of $P_dP_d$ is relatively large.

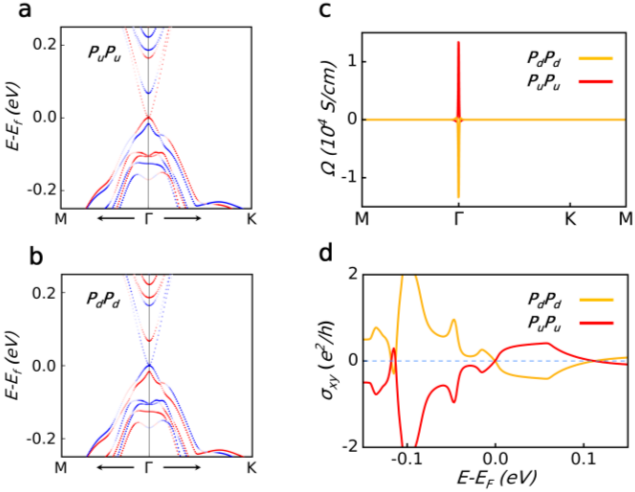

Fig. 6 (**a**, **b**) Energy bands of the *FP*-MBT films with interlayer electric polarization in both top and bottom interface and the pointing-up (TBIP-$P_uP_u$) (a) or -down (TBIP-$P_dP_d$) (b). The red region represents dominating spin-*up* energy bands, while the blue region represents the spin-*down* one. (**c**) Berry curvature of the TBIP-$P_uP_u$/-$P_dP_d$ along the high symmetric directions at Fermi energy. (**d**) Anomalous Hall conductivity as a function of electron energy for TBIP-$P_uP_u$ and -$P_dP_d$.

Apart from the two nearest-neighbor polar interfaces mentioned above, we also analyze two polar interfaces that are separated by a nonpolar interface. One of these polar interfaces exists within layer1 and layer2, while the other one locates between layer3 and layer4. The polarization of the two interfaces points along the same direction, either up or down, and it is named *TBIP-$P_uP_u$*, *TBIP-$P_dP_d$*, respectively. These is a nopolar interface between top and bottom electric polar interface. These polar structures could be achieved through contemporarily sliding layer1 and layer4 along the **t**$_{//}$ or -**t**$_{//}$ direction. Our calculations show that for both *TBIP-$P_uP_u$* and *TBIP-$P_dP_d$* state, they maintain *A*-type antiferromagnetic order of the bulk-like four-layer MnBi$_2$Te$_4$ film. Our symmetry analysis shows that, even for each structure, the $\hat{P}\hat{T}$ has been broken because of the interlayer polarization, there is a $\hat{P}\hat{T}$ symmetry correlation between *TBIP-$P_uP_u$* and *TBIP-$P_dP_d$* polar structures. We can clearly see from Fig. 6(a) and 6(b) that with the reversal of the interlayer polarization direction, the spin states that dominate in the energy bands undergo a shift, which is manifested as a change in the color of the corresponding position. These symmetry relations also result in distinctive topological transport properties between *TBIP-$P_uP_u$* and *TBIP-$P_dP_d$*. The results in Fig. 6(c) clearly shows that, the broken of $\hat{P}\hat{T}$ symmetry in either *TBIP-$P_uP_u$* or *TBIP-$P_dP_d$* leads to the non-vanishing berry curvature, the peak is approximately $\pm 1.2 \times 10^4$ S/cm at the $\Gamma$ point. Moreover, the sign of the berry curvature is totally reversed with the reversal of the polarization direction because of the $\hat{P}\hat{T}$ symmetry connection ($\hat{P}\hat{T}\Omega(k) = -\Omega(k)$). Furthermore, the calculated AHC($\sigma_{xy}$) in Fig. 6(d) clearly show that, because of the $\hat{P}\hat{T}$ symmetry relation between *TBIP-$P_uP_u$* and -$P_dP_d$ states, the direction of AHC in *TBIP FP-MBT* is totally determined by the direction of interlayer polarization. In the end, we also analyze four-layer MnBi$_2$Te$_4$ film with three polarization interfaces (*AIP-$P_u$*, *AIP-$P_d$*), the specific atomic structure is shown in Fig. S6, the higher polarization acts strongly on the bending of the band edges, leading to the closure of bandgap in both *AIP-$P_u$* and *AIP-$P_d$*, and finally only AHC is expected in fully polar structures.

## IV. SUMMARY

We have successfully implemented switchable interlayer polarization in MnBi$_2$Te$_4$ films with a four-layer Van der Waals structure. Based on symmetry analysis and topological transport property calculations, we proposed that anomalous

Hall effect (AHE), and even quantum anomalous Hall effect (QAHE) could be efficiently induced and modulated by interlayer polarization in four-layer MBT. The polar interface in four-layer MnBi$_2$Te$_4$ introduces a polarization-locked magnetic order. The generation of a ferromagnetic interface and breaking of the intrinsic $\hat{P}\hat{T}$ symmetry in four-layer MnBi$_2$Te$_4$ collectively induce novel topological transport properties that cannot be in bulk state. Switching the polarization at the top or bottom interfaces has the capability to induce and modify the AHE and QAHE in a four-layer MnBi$_2$Te$_4$. Switching the polarization at the middle interfaces results in a reversal of the direction of the quantum anomalous hall current. In addition, the interlayer polarization of the top and bottom interfaces may be in the same direction, and reversing the polarization direction also leads to a reversal of the anomalous hall current. These diverse quantum states, which are determined by interlayer polarization, can be readily achieved in experiments using vdW assembly. The exploration of diverse quantum states in MnBi$_2$Te$_4$ contributes significantly. Moreover, it holds potential for future applications in electronic or spintronic devices.

[1] L. J. Du, T. Hasan, A. Castellanos-Gomez, G. B. Liu, Y. G. Yao, C. N. Lau, and Z. P. Sun, Nature Reviews Physics **3**, 193 (2021).
[2] C. Z. Chang *et al.*, Science **340**, 167 (2013).
[3] T. Taniguchi, Physical Review B **94**, 174440 (2016).
[4] J. H. Li, Y. Li, S. Q. Du, Z. Wang, B. L. Gu, S. C. Zhang, K. He, W. H. Duan, and Y. Xu, Science Advances **5**, eaaw5685 (2019).
[5] Y. J. Deng, Y. J. Yu, M. Z. Shi, Z. X. Guo, Z. H. Xu, J. Wang, X. H. Chen, and Y. B. Zhang, Science **367**, 895 (2020).
[6] L. Smejkal, R. González-Hernández, T. Jungwirth, and J. Sinova, Science Advances **6**, eaaz8809 (2020).
[7] T. Nan *et al.*, Nature Communications **11**, 4671 (2020).
[8] G. Gurung, D. F. Shao, and E. Y. Tsymbal, Physical Review Materials **5**, 124411 (2021).
[9] Z. Wang *et al.*, Nature Nanotechnology **13**, 554 (2018).
[10] J. Q. Cai *et al.*, Nature Communications **13**, 1668 (2022).
[11] L. Li and M. H. Wu, Acs Nano **11**, 6382 (2017).
[12] F. R. Sui, M. Jin, Y. Y. Zhang, R. J. Qi, Y. N. Wu, R. Huang, F. Y. Yue, and J. H. Chu, Nature Communications **14** (2023).
[13] P. Meng *et al.*, Nature Communications **13**, 7696 (2022).
[14] S. Deb *et al.*, Nature **612**, 465 (2022).
[15] Z. Y. Fei, W. J. Zhao, T. A. Palomaki, B. S. Sun, M. K. Miller, Z. Y. Zhao, J. Q. Yan, X. D. Xu, and D. H. Cobden, Nature **560**, 336 (2018).
[16] Q. Yang, M. H. Wu, and J. Li, Journal of Physical Chemistry Letters **9**, 7160 (2018).
[17] K. Yasuda, X. R. Wang, K. Watanabe, T. Taniguchi, and P. Jarillo-Herrero, Science **372**, 1458 (2021).
[18] M. V. Stern *et al.*, Science **372**, 1462 (2021).
[19] O. Dahl, J. K. Grepstad, and T. Tybell, Journal of Applied Physics **106**, 084104 (2009).
[20] T. F. Cao, G. D. Ren, D. F. Shao, E. Y. Tsymbal, and R. Mishra, Physical Review Materials **7**, 044412 (2023).
[21] Z. Y. Hao, A. M. Zimmerman, P. Ledwith, E. Khalaf, D. H. Najafabadi, K. Watanabe, T. Taniguchi, A. Vishwanath, and P. Kim, Science **371**, 1133 (2021).
[22] Y. R. Zhang *et al.*, Science **377**, 1538 (2022).
[23] X. R. Wang, K. Yasuda, Y. Zhang, S. Liu, K. Watanabe, T. Taniguchi, J. Hone, L. Fu, and P. Jarillo-Herrero, Nature Nanotechnology **17**, 367 (2022).
[24] Y. Gong *et al.*, Chinese Physics Letters **36**, 076801 (2019).
[25] M. M. Otrokov *et al.*, Nature **576**, 416 (2019).
[26] C. Liu *et al.*, Nature Materials **19**, 522 (2020).
[27] A. Y. Gao *et al.*, Nature **595**, 521 (2021).
[28] M. M. Otrokov *et al.*, Physical Review Letters **122**, 107202 (2019).

[29] W. Y. Lin *et al.*, Nature Communications **13**, 7714 (2022).

[30] D. Q. Zhang, M. J. Shi, T. S. Zhu, D. Y. Xing, H. J. Zhang, and J. Wang, Physical Review Letters **122**, 206401 (2019).

[31] T. F. Cao, D. F. Shao, K. Huang, G. Gurung, and E. Y. Tsymbal, Nano Letters **23**, 3781 (2023).

[32] G. Kresse and J. Furthmuller, Physical Review B **54**, 11169 (1996).

[33] J. P. Perdew, K. Burke, and M. Ernzerhof, Physical Review Letters **78**, 1396 (1997).

[34] V. I. Anisimov, J. Zaanen, and O. K. Andersen, Physical Review B **44**, 943 (1991).

[35] S. L. Dudarev, G. A. Botton, S. Y. Savrasov, C. J. Humphreys, and A. P. Sutton, Physical Review B **57**, 1505 (1998).

[36] D. F. Shao, J. Ding, G. Gurung, S. H. Zhang, and E. Y. Tsymbal, Physical Review Applied **15**, 024057 (2021).

[37] P. Li, J. Y. Yu, Y. Wang, and W. D. Luo, Physical Review B **103**, 155118 (2021).

[38] S. Grimme, J. Antony, S. Ehrlich, and H. Krieg, Journal of Chemical Physics **132**, 154104 (2010).

[39] J. Neugebauer and M. Scheffler, Physical Review B **46**, 16067 (1992).

[40] G. Pizzi *et al.*, Journal of Physics-Condensed Matter **32**, 165902 (2020).

[41] Q. S. Wu, S. N. Zhang, H. F. Song, M. Troyer, and A. A. Soluyanov, Computer Physics Communications **224**, 405 (2018).

[42] R. Peng, T. Zhang, Z. L. He, Q. Wu, Y. Dai, B. B. Huang, and Y. D. Ma, Physical Review B **107**, 085411 (2023).